\documentstyle[12pt]{article}

\def\thefootnote{\fnsymbol{footnote}}
\catcode`\@=11
\begin{document}
\begin{titlepage}
\begin{center}
\hfill CERN-TH-6784/93\\
\vskip 1.in
{\large \bf
Symmetry Aspects and Finite-Size Scaling \\of \\Quantum Hall Fluids\footnote{
Based on the talks presented at the Conference on {\it Condensed Matter and
High-Energy Physics}, Chia Laguna (Sardinia), September 1992,
to appear in Nucl. Phys. B (Proc. Suppl.), L. Alvarez-Gaum\'e et al. eds..}}
\vskip 0.8in
Andrea CAPPELLI${}^{ a}$\footnote{
On leave from INFN, Largo E. Fermi 2, I-50125 Firenze, Italy.},
Gerald V. DUNNE${}^{ b}$,\\
Carlo A. TRUGENBERGER${}^{a}$ and Guillermo R. ZEMBA${}^{a}$
\\[.2in]
{\em ${}^a$ Theory Division, CERN, 1211 Geneva 23, Switzerland} \\
{\em ${}^b$ Dept. of Physics, Univ. of Connecticut, 2152 Hillside Road,
Storrs, CT 06268 USA}\\
\end{center}
\vskip .5in

\begin{abstract}
The exactness and universality observed in the quantum Hall effect suggests
the existence of a symmetry principle underlying Laughlin's theory.
We review the role played by the infinite
$W_{\infty }$ and conformal algebras as dynamical symmetries of incompressible
quantum fluids and show how they predict universal finite-size effects in the
excitation spectrum.
\end{abstract}
\vfill
CERN-TH-6784/93\hfill\\
January 1993\hfill\\
\end{titlepage}
\end{document}

\def\fileversion{v2.0}
\def\filedate{5 May 1992}

\typeout{Document style option `espcrc2' \fileversion \space\space
         <\filedate>}

\oddsidemargin  -4mm              
\evensidemargin  4mm              

\topmargin      16mm              
\headheight      0mm              
\headsep         0mm              
\footskip        30pt             

\textheight 202mm                 
\textwidth 160mm                  

\columnsep 10mm                   
\columnseprule 0pt                

\parskip 0pt                      
\parindent 1em                    

\newdimen\@bls                    
\@bls=\baselineskip               
\advance\@bls -1ex                
\newdimen\@eps                    %
\@eps=0.0001pt                    

\def\section{\@startsection{section}{1}{\z@}
  {2\@bls plus 0.5\@bls}{1\@bls}{\normalsize\bf}}
\def\subsection{\@startsection{subsection}{2}{\z@}
  {1\@bls plus 0.25\@bls}{\@eps}{\normalsize\bf}}
\def\subsubsection{\@startsection{subsubsection}{3}{\z@}
  {1\@bls plus 0.25\@bls}{\@eps}{\normalsize\bf}}
\def\paragraph{\@startsection{paragraph}{4}{\parindent}
  {1\@bls plus 0.25\@bls}{0.5em}{\normalsize\bf}}
\def\subparagraph{\@startsection{subparagraph}{4}{\parindent}
  {1\@bls plus 0.25\@bls}{0.5em}{\normalsize\bf}}

\def\@sect#1#2#3#4#5#6[#7]#8{\ifnum #2>\c@secnumdepth
  \def\@svsec{}\else
  \refstepcounter{#1}\edef\@svsec{\csname the#1\endcsname.\hskip0.5em}\fi
  \@tempskipa #5\relax
  \ifdim \@tempskipa>\z@
    \begingroup
      #6\relax
      \@hangfrom{\hskip #3\relax\@svsec}{\interlinepenalty \@M #8\par}%
    \endgroup
    \csname #1mark\endcsname{#7}\addcontentsline
      {toc}{#1}{\ifnum #2>\c@secnumdepth \else
        \protect\numberline{\csname the#1\endcsname}\fi #7}%
  \else
    \def\@svsechd{#6\hskip #3\@svsec #8\csname #1mark\endcsname
      {#7}\addcontentsline{toc}{#1}{\ifnum #2>\c@secnumdepth \else
        \protect\numberline{\csname the#1\endcsname}\fi #7}}%
  \fi \@xsect{#5}}

\long\def\@makefigurecaption#1#2{\vskip 10mm #1. #2\par}

\long\def\@maketablecaption#1#2{\hbox to \hsize{\parbox[t]{\hsize}
  {#1 \\ #2}}\vskip 0.3ex}

\def\fnum@figure{Figure \thefigure}
\def\figure{\let\@makecaption\@makefigurecaption \@float{figure}}
\@namedef{figure*}{\let\@makecaption\@makefigurecaption \@dblfloat{figure}}

\def\table{\let\@makecaption\@maketablecaption \@float{table}}
\@namedef{table*}{\let\@makecaption\@maketablecaption \@dblfloat{table}}

\floatsep 10mm plus 4pt minus 4pt 
\textfloatsep=\floatsep           
\intextsep=\floatsep              

\long\def\@makefntext#1{\parindent 1em\noindent\hbox{${}^{\@thefnmark}$}#1}

\mathindent=0em

\def\maketitle{\begingroup        
    \def\thefootnote{\fnsymbol{footnote}}%
    \newpage \global\@topnum\z@
    \@maketitle \thispagestyle{plain}\@thanks
  \endgroup
  \let\maketitle\relax \let\@maketitle\relax
  \gdef\@thanks{}\let\thanks\relax
  \gdef\@address{}\gdef\@author{}\gdef\@title{}\let\address\relax}

\def\justify@on{\let\\=\@normalcr
  \leftskip\z@ \@rightskip\z@ \rightskip\@rightskip}

\newbox\fm@box                    

\def\@maketitle{
  \global\setbox\fm@box=\vbox\bgroup
    \raggedright                  
    \hyphenpenalty\@M             
    {\Large \@title \par}         
    \vskip\@bls                   
    {\normalsize                  
     \@author \par}               
    \vskip\@bls                   
    \@address                     
  \egroup
  \twocolumn[
    \unvbox\fm@box                
    \vskip\@bls                   
    \unvbox\abstract@box          
    \vskip 2pc]}                  

\newcounter{address}
\def\theaddress{\alph{address}}
\def\@makeadmark#1{\hbox{$^{\rm #1}$}}

\def\address#1{\addressmark\begingroup
  \xdef\@tempa{\theaddress} \let\\=\relax
  \def\protect{\noexpand\protect\noexpand}\xdef\@address{\@address
  \protect\addresstext{\@tempa}{#1}}\endgroup}
\def\@address{}

\def\addressmark{\stepcounter{address}%
  \xdef\@tempa{\theaddress}\@makeadmark{\@tempa}}

\def\addresstext#1#2{\leavevmode \begingroup
  \raggedright \hyphenpenalty\@M \@makeadmark{#1}#2\par \endgroup
  \vskip\@bls}

\newbox\abstract@box              

\def\abstract{%
  \global\setbox\abstract@box=\vbox\bgroup
  \small\rm
  \ignorespaces}
\def\endabstract{\par \egroup}

\def\thebibliography#1{\section*{REFERENCES}\list{\arabic{enumi}}
  {\settowidth\labelwidth{#1}\leftmargin=1.67em
   \labelsep\leftmargin \advance\labelsep-\labelwidth
   \itemsep\z@ \parsep\z@
   \usecounter{enumi}}\def\makelabel##1{\rlap{##1}\hss}%
   \def\newblock{\hskip 0.11em plus 0.33em minus -0.07em}
   \sloppy \clubpenalty=4000 \widowpenalty=4000 \sfcode`\.=1000\relax}

\newcount\@tempcntc
\def\@citex[#1]#2{\if@filesw\immediate\write\@auxout{\string\citation{#2}}\fi
  \@tempcnta\z@\@tempcntb\m@ne\def\@citea{}\@cite{\@for\@citeb:=#2\do
    {\@ifundefined
       {b@\@citeb}{\@citeo\@tempcntb\m@ne\@citea
        \def\@citea{,\penalty\@m\ }{\bf ?}\@warning
       {Citation `\@citeb' on page \thepage \space undefined}}%
    {\setbox\z@\hbox{\global\@tempcntc0\csname b@\@citeb\endcsname\relax}%
     \ifnum\@tempcntc=\z@ \@citeo\@tempcntb\m@ne
       \@citea\def\@citea{,\penalty\@m\ }
       \hbox{\csname b@\@citeb\endcsname}%
     \else
      \advance\@tempcntb\@ne
      \ifnum\@tempcntb=\@tempcntc
      \else\advance\@tempcntb\m@ne\@citeo
      \@tempcnta\@tempcntc\@tempcntb\@tempcntc\fi\fi}}\@citeo}{#1}}

\def\@citeo{\ifnum\@tempcnta>\@tempcntb\else\@citea
  \def\@citea{,\penalty\@m\ }%
  \ifnum\@tempcnta=\@tempcntb\the\@tempcnta\else
   {\advance\@tempcnta\@ne\ifnum\@tempcnta=\@tempcntb \else \def\@citea{--}\fi
    \advance\@tempcnta\m@ne\the\@tempcnta\@citea\the\@tempcntb}\fi\fi}

\sloppy                         
\emergencystretch=1pc           
\flushbottom                    



\documentstyle[twoside,fleqn,espcrc2]{article}

\def\beq{\begin{equation}}
\def\eeq{\end{equation}}
\def\barr{\begin{eqnarray}}
\def\earr{\end{eqnarray}}

\newcommand{\ttbs}{\char'134}
\newcommand{\AmS}{{\protect\the\textfont2
  A\kern-.1667em\lower.5ex\hbox{M}\kern-.125emS}}

\hyphenation{financial created another}

\title{Symmetry aspects and finite-size scaling of quantum Hall fluids}

\author{Andrea Cappelli\address{Theory Division, CERN, 1211 Geneva 23,
        Switzerland}\thanks{On leave from INFN, Largo E. Fermi 2,
        I-50125 Firenze, Italy.},
        Gerald V. Dunne\address{Dept. of Physics, University of Connecticut,
        2152 Hillside Road, Storrs, CT 06268 USA},
        Carlo A. Trugenberger ${}^a$ and Guillermo R. Zemba ${}^a$}

\begin{document}

\begin{abstract}
The exactness and universality observed in the quantum Hall effect suggests
the existence of a symmetry principle underlying Laughlin's theory.
We review the role played by the infinite
$W_{\infty }$ and conformal algebras as dynamical symmetries of incompressible
quantum fluids and show how they predict universal finite-size effects in the
excitation spectrum.
\end{abstract}

\maketitle

\section{INTRODUCTION}

The quantum Hall effect \cite{Pra1}
provides fascinating examples of {\it quantum
fluids}. At low temperatures, interacting planar electrons in high
magnetic fields $B$ have strong quantum correlations which lead to
collective motion and macroscopic quantum effects. These find their
experimental evidence in a discrete series of plateaus at rational
values of the Hall conductivity:
\barr
\sigma _{xy}&=&{ e^2\over h}\ \nu\ ,\nonumber\\
\nu &=& 1, {1\over 3}, {1\over 5},{2\over 7},\dots,2,\dots .
\label{0}\earr
Corresponding to these plateaus the longitudinal
conductivity $\sigma _{xx}$ vanishes.
The same plateaus are observed in several materials, signalling
{\it universality}.
Another and more important experimental result is
the remarkable {\it exactness} of these rational values of $\nu $;
the experimental error is $\Delta\nu=10^{-8}$ for integer $\nu$.
Moreover, upon assuming exact integer values one obtains a measure
of the fine-structure constant $\alpha=e^2/\hbar c$ consistent with
the best result from Quantum Electrodynamics and equally accurate
\cite{Kino}.

The current understanding of the quantum Hall effect is based
on the seminal work of Laughlin \cite{Lau1}. The main idea is the
existence of {\it incompressible quantum fluids} at specific rational
values $\bar\rho =\nu B/2\pi $ of the electron density. These are very stable,
macroscopical quantum states with uniform density and an energy gap.
Incompressibility
accounts for the lack of low-lying conduction modes, which causes
$\sigma _{xx}$ to vanish, while the Hall conduction is realized as an
overall rigid motion of the uniform droplet, which gives eq.(\ref{0}).

Laughlin's theory comprises a body of theoretical arguments substantiated
by extensive numerical computations. Combined with the hierarchical
constructions of Haldane and Halperin \cite{Hal1} and of Jain \cite{Jai1} it
accounts for all observed fractions and predicts also the quantum
numbers of excitations.

While Laughlin's theory is very successful, the observed exactness and
universality calls for a {\it fundamental principle} underlying it.
Exact rational numbers, universality and a simple structure of the excitation
spectrum usually indicate that dynamics is dominated by {\it symmetry}.
The {\it effective field-theory approach} can be useful in this respect.
One introduces an effective field theory with the relevant symmetry
and describes the universal, long-range properties of the system.
For example, the quantum numbers of pions and low-lying hadrons
and their low-energy interactions can be explained by spontaneously
broken chiral symmetry, without knowing the dynamics of
confinement \footnote{
Spontaneous symmetry breaking in superconductivity is discussed in the
same spirit by Fubini and Molinari in this volume.}.

In two dimensions, the more powerful {\it infinite conformal symmetry}
\cite{Gin1} is sufficient to solve the field theory exactly.
In fact, this symmetry determines the complete excitation spectrum,
which possesses exact fractional levels.
This has
prompted a number of authors \cite{Fub1} to establish a formal analogy
between the quantum Hall effect and conformal invariant field theory (CFT).
This analogy was further extended in terms of Chern-Simons gauge theory
in \cite{Fro1}.

Here we discuss the infinite symmetry underlying Laughlin's approach.
Furthermore, we derive an effective conformal field theory for the
excitations at the edge of the Hall sample.
This theory accounts for the universality and exactness
of the observed filling fractions $\nu $ and gives the quantum numbers
of excitations and the structure of their spectrum.
Our approach should not be considered as an alternative to Laughlin's
theory but complementary to it.
We cannot compute the exact mass gaps because we cannot solve the dynamics
of electrons (as in previous examples of effective field theories),
nor can we describe the effect of impurities.
However, the symmetry principle explains the stability of the incompressible
fluid, the key property for understanding the observed exact Hall
conductivities.
The material presented here is based
on two recent papers \cite{Cap1}, \cite{Cap2} and represents only a
stage of work in progress.

The basic idea is simple: at the classical level, a two-dimensional
incompressible fluid forms droplets of constant area. Different
configurations are related by area-preserving diffeomorphisms.
The generators of these transformations satisfy an infinite algebra called
$w_{\infty }$. Two-dimensional incompressible {\it quantum}
fluids can thus be expected to be characterized by
representations of its quantum counterpart
$W_{\infty }$ \cite{She1}.
This is indeed the picture that emerges from the analysis
in section three. We stress that $W_{\infty }$ is a {\it dynamical symmetry},
like conformal symmetry in two dimensions: the Hamiltonian is an element
of the algebra, the ground state satisfies highest-weight conditions and
excitations are classified by irreducible representations of the
algebra \cite{Gin1}.

In sections four and five we show how a CFT with central charge $c=1$
emerges when one consider the dynamics of
excitations at the edge of the quantum fluid droplet. While the mass
gaps are beyond the scope of our analysis, CFT predicts that their finite-size
corrections are universal numbers related to the charge of the excitations.
This provides a new test of Laughlin's theory based on subleading effects.

\section{THE INCOMPRESSIBLE QUANTUM FLUIDS}

We consider $N$ spin-polarized planar electrons moving in a uniform
magnetic field and interacting via short-range two-body potentials.
We also include in the Hamiltonian a confining potential
to simulate the finite size of the sample.
In units $\hbar =1$, $c=1$, the total
Hamiltonian is thus given by
\barr
H &=& H_B+H_I +H_C \ ,\label{1}\\
H_B &=&- \sum_{i=1}^N {1\over 2M} \left( \vec{\nabla }_i -ie\vec{A}_i
        \right)^2 -{eBN\over 2M} \ ,\label{2}\\
H_I &=& \sum_{k=1\  {\rm odd}}^{m-2} \ V_k \sum_{i<j=1}^N \triangle ^{k}
      \delta ^2 (\vec{x}_i-\vec{x}_j) \ ,\label{3}\\
H_C &=& \sum_{i=1}^N V_C(\vec{x}_i) \ .\label{4}
\earr

The Hamiltonian $H_B$ describes the interaction of the electrons with the
external magnetic field including the Pauli interaction.
Since we shall deal only with circular geometries
we choose to use the symmetric gauge $\vec{A}={B\over 2}(-y, x)$ for the
external vector potential.
The external magnetic field introduces a unit of length,
the magnetic length $\ell =\sqrt{2/eB}$. In the following we will
make extensive use of the complex notation $z=x+iy$, $\bar z=x-iy$,
$\partial =\partial /\partial z$, $\bar \partial =\partial /\partial \bar z$.
By introducing two commuting harmonic oscillator operators
\barr
d &=& {z\over 2\ell}+\ell \bar \partial \ ,\ d^{\dag }={\bar z\over 2\ell}
-\ell \partial \ ,\nonumber \\
c &=& {\bar z\over 2\ell}+\ell \partial \ ,\ c^{\dag }={z\over 2\ell}
-\ell \bar\partial \ ,
\label{5}
\earr
the one-body Hamiltonian and (canonical) angular momentum
$J=-i\vec{x}\wedge \vec{\nabla}$ can be rewritten as
\barr
H &=& \omega \ d^{\dag}d \ ,\nonumber \\
J &=& c^{\dag}c-d^{\dag}d \ ,\label{6}
\earr
where $\omega =eB/M$ is the cyclotron frequency. Since the operators $c$
and $d$ commute, the spectrum consists of infinitely degenerate levels of
energy $\epsilon_n=\omega n$; these are called the Landau levels. The
degenerate states in one Landau level are characterized by the angular
momentum eigenvalue $l$.
For the physical problems we shall address it is a good approximation
to truncate the theory to the {\it first Landau level} \footnote{
Actually, some of the following results hold for the full theory too
\cite{Cap1}.}.
Wave functions of this level satisfy $d\psi =0$; a complete orthonormal
basis is given by
\beq
\psi _l(z, \bar z)={1\over \ell \sqrt{\pi}} {1\over \sqrt{l!}} \left(
{z\over \ell}\right)^l {\rm exp}\left(- {|z|^2\over 2\ell ^2}\right)
\ .\label{7}
\eeq
Note that the radial part of these wave functions is sharply peaked
around a radius $r_l=\ell \sqrt{l}$.

In second quantization $H_B$ takes the form
\beq
H_B={1\over 2M}\int d^2\vec{x} \left[
\left(\vec{D}\Psi \right) ^{\dag}
\left(\vec{D}\Psi \right) - eB\rho \right] \ ,\label{8}
\eeq
with $\vec{D}=\vec{\nabla}-ie\vec{A}$ the covariant derivative and
$\rho =\Psi ^{\dag }\Psi $ the particle-number density. When restricted
to the first Landau level, the field operator $\Psi $ possesses an
expansion in terms of the single-particle angular momentum eigenstates
(\ref{7}):
\beq
\Psi (\vec{x}, t)=\sum _{l=0}^{\infty }a_l \psi _l(\vec{x}) \ .
\label{9}
\eeq
The coefficients are fermionic Fock annihilators satisfying
\beq
\left\{a_k,a_l^{\dag }\right\}=\delta _{kl} \ ,
\label{10}
\eeq
with all other anticommutators vanishing.

In the many-body problem of $N$ electrons in an external magnetic field
there are two fundamental scales: one is the magnetic length discussed
above, which is the natural ultraviolet cutoff; the other is the average
dimension of the area occupied by the electrons and provides the natural
infrared cutoff. These two scales are tied together by the particle number
$N=\bar\rho\ Area$ as follows:
\beq
\bar \rho ={\nu \over 2\pi }eB \ ,
\label{11}
\eeq
where $\bar \rho $ is the average particle density and therefore $1/\nu $
represents the magnetic flux per particle in quantum units. The quantity
$\nu $, called the {\it filling fraction}, is a dimensionless measure of
the electron density. When only states in the first Landau level are
occupied, an equivalent definition of the filling fraction is
\beq
\nu={N(N-1)\over 2J} \ ,
\label{12}
\eeq
i.e. the ratio of the minimal angular momentum allowed by Fermi statistics
to the actual angular momentum of the state.

The Hamiltonian $H_I$ in (\ref{3}) describes the electron-electron
interactions. It is a short hand notation \cite{Tru1} for Haldane's
short-range potentials \cite{Hal1}, which are constructed
as follows. When projected (P) onto the first Landau level, every
two-body potential $V(|\vec{x}_1-\vec{x}_2|)$ admits a
``short-distance expansion'' \cite{Gir1}
\beq
PV(|\vec{x}_1-\vec{x}_2|)P=\sum _{k {\rm odd}} V_k
P_k(\vec{x}_1 - \vec{x}_2) \ ,
\label{13}
\eeq
where $P_k(\vec{x}_1 - \vec{x}_2)$ is the projector on states for which
particles $1$ and $2$ have relative angular momentum $k$.
These projectors can be represented in matrix elements by
$\triangle^k \delta^2(\vec{x}_1-\vec{x}_2)$.
The $V_k$ are "Fourier coefficients" depending on the original potential $V$.
Haldane's potentials consist of finite sums of $P_k$ in (\ref{13}) up to
the odd number $(m-2)$ with positive coefficients $V_k=O(1)$ (their exact
values
will not be relevant here).

\begin{figure*}[t]
  \vspace{7cm}
  \caption{The density profile in units of $1/\pi\ell^2$ for the first
Landau level filled up to $L=50$ as a function of $r/\ell$.}
\end{figure*}

Let us first analyze the situation in which all $V_k$ vanish, i.e.
non-interacting electrons. In this case we have the freedom of filling
all angular momentum states of the first Landau level up to (and
including) a maximal angular momentum $L=N-1$, and to construct thus
a ground state of filling fraction $\nu =1$. This state is described by
\beq
\psi_1(z_1\dots z_N)=\prod _{i<j}(z_i-z_j) \ \exp\left(
-\sum _{i=1}^N{|z_i|^2\over 2\ell ^2} \right)
\label{14}
\eeq
in first quantization, and by
\beq
|\Omega \rangle _1=a_0^{\dag }\dots a_L^{\dag }|0\rangle
\label{15}
\eeq
in second quantization. Since the single-particle angular momentum states
are peaked around radii $r_l=\ell \sqrt{l}$, the above state consists of a
circular droplet of radius approximately given by $R\simeq \ell \sqrt{L}$.
For $r < R$ the density of the droplet is uniform: $\rho =1/\pi \ell ^2$
(see Figure 1).
This configuration is clearly {\it incompressible}: a compression of the
droplet would lower its total angular momentum and face an energy gap $\omega
$,
since at least one electron would have to be promoted to the next Landau level.
However, decompressions, i.e. transitions to states with higher angular
momentum cost no energy due to the degeneracy of the Landau level. States
with angular momentum $J=N(N-1)/2+\Delta J$ are generically
described in first quantization by
\beq
\psi _{\Delta J}(z_1\dots z_N)=P_{\Delta J}(z_1\dots z_N)\psi _1
(z_1\dots z_N)
\label{16}
\eeq
where $P_{\Delta J}$ is a symmetric homogeneous polynomial of
degree $\Delta J$. In second quantization they correspond to transitions
of electrons from states of single-particle angular momentum smaller than
$L$ ("inside the droplet") to states with single-particle angular momentum
bigger than $L$ ("outside the droplet"). For $\Delta J =O(1)$ these
transitions clearly affect only the edge of the droplet, leaving its
interior (bulk) unchanged.

Let us now turn on the interaction $H_I$. The effect of this truncated
interaction is to divide the $N$-particle Hilbert space into two sub-bands.
All states in which every couple of electrons has relative angular momentum
at least equal to $m$ remain zero energy eigenstates; all other states,
containing at least one couple with relative angular momentum less than $m$,
acquire a positive energy depending on the coefficients $V_k$ in (\ref{3}).
The zero energy eigenspace is thus parametrized as
\beq
\psi _{\Delta J}(z_1\dots z_N)=P_{\Delta J}(z_1\dots z_N)\psi _m
(z_1\dots z_N)
\label{17}
\eeq
where
\beq
\psi_m(z_1\dots z_N)=\prod_{i<j} (z_i-z_j)^{m-1}
\psi_1(z_1\dots z_N)
\label{18}
\eeq
is the {\it Laughlin wave function} at level $m$. This wave function
describes a state with angular momentum $J=mN(N-1)/2$, i.e. filling
fraction $\nu =1/m$. Thus, Haldane's construction provides an energy
gap for all states with filling $\nu >1/m$, which is tantamount to the
{\it incompressibility} of the Laughlin states $\psi _m$.

The last term $H_C$ in the Hamiltonian (\ref{1}) describes the confining
potential which keeps the electrons together. In a real sample of finite
size, say a disk of radius $R$, the Landau levels are deformed near the edge.
In particular, the energy increases for increasing angular
momentum so that the electrons are confined \cite{Halp}.
Note that, in the disk geometry, the filling fraction can be changed
by varying $B$, according to eq.(\ref{11}).

Here we replace the disk with the infinite plane plus a confining
potential, which we take to be harmonic:
\beq
V_C\left(|\vec{x}|\right)={1\over 2} \lambda |\vec{x}|^2 \ , \ \ \
\lambda=O\left({\bar\rho\over R}\right) \ .
\label{19}
\eeq
This gives a linear spectrum of states in the first Landau level:
\barr
H_B+H_C &=& \alpha J \nonumber \\
\alpha &=& {1\over 2}\sqrt{\omega ^2 +{4\lambda \over M}}-{\omega \over 2}
\ ,\label{20}
\earr
(we have omitted an irrelevant overall constant).

There are two important physical points associated with the confining
potential. First, as a consequence of this
potential also decompressions of the states $\psi _m$ (transitions
to higher angular momentum) cost energy and $\psi _m$ becomes the unique,
non-degenerate ground state of $H$. Secondly, the physical values of
$\lambda =O(\bar\rho /R)$ give \hbox{$\alpha =O(1/R)$}, in the relevant regime
$R\omega \gg 1$. Thus, the states $\psi _{\Delta J}$
in eq.(\ref{17}) with $\Delta J =O(1)$ become {\it gapless
edge excitations} \cite{Wen1}; these are the low-lying
excitations about the incompressible Laughlin ground states
in geometries with a boundary.

Equation (\ref{20}) with $\alpha=O(1/R)$ represents the leading term in the
$R\to\infty$ expansion of the lifting of the Landau level degeneracy
by a generic confining potential.
Subleading effects originates, for example, from the interaction
of electrons with the neutralizing background of the lattice.
This can be derived by computing the energy of the plasma oscillation
corresponding to an elementary edge excitations of an electron moved up
in angular momentum. This results in a $O(1/R^2)$ effect.

Note that in the infinite plane, the {\it pressure} $\alpha$ controls the
angular momentum of the ground state $\langle J\rangle$ (its conjugate
variable by eq.(\ref{20})) and allows to vary the filling fraction
according to eq.(\ref{12}).

Having discussed in detail the physical model we shall investigate,
we should briefly mention the role of impurities and localization.
The physics of localization is crucial for understanding
the formation of the Hall plateaus \cite{Pra3} and
the transitions between plateaus \cite{Prui}.
At the center of a given plateau, the physics of the incompressible fluid
and its excitations is independent of impurities and
can be described in the idealized limit of pure samples \cite{Halp}.
Since the analysis of the latter properties is our main goal,
we do not consider impurities in our effective theory approach.

Experimentally, the Hall plateaus get narrower for both very small and
very large concentrations of impurities.
A large concentration clearly destroys the incompressible fluid,
while a small concentration does not;
it give rise to localized states with energies in between the Landau bands.
When the filling is slightly off the incompressible value,
charged excitations are produced.
These populate the localized states and do not contribute to the Hall
conduction, which thus remains constant at the incompressible value
\cite{Pra3}.
Therefore, while impurities are crucial for the formation of the Hall
plateaus (i.e. for the experimental observation of incompressible ground
states), they do not determine the properties of these
states and their excitations.

{}From now on we choose units such that the magnetic length
$\ell =1$, so that $eB=2$. For simplicity of presentation we also
choose $M=e=1$ in these units.

\section{DYNAMICAL $W_\infty$ SYMMETRY IN THE BULK}

In this section we review the symmetry properties of the incompressible
Laughlin fluids from the point of view of the $(2+1)$-dimensional
non-relativistic theory {\it (symmetries in the bulk)}.
We first consider the $\nu=1$ ground state of non-interacting electrons
and then move on to discuss the $\nu=1/m$ fluids for generic odd $m$.
This will lead to a {\it geometric and algebraic characterization of
incompressibility}.

The infinite degeneracy of the Landau levels of non-interacting electrons
is due to the vanishing commutators between the Hamiltonian $H_B$
and the generators $c_i,c^{\dag}_i$ of magnetic translations,
see (\ref{5}) and (\ref{6}).
In the absence of the confining potential, conserved charges can
therefore be constructed by taking arbitrary combinations of
powers of the operators $c_i$ and $c^{\dag}_i$:
\beq
\begin{array}{l}{\displaystyle
{\cal L}_{n,m} \equiv \sum_{i=1}^{N}  c_i^{\dag n+1} c_i^{m+1}\ ,
\quad n,m\ge -1\ ,} \\
{\displaystyle
{[{\cal L}_{n,m},H_B]} =0\ .}
\end{array}
\label{21}\eeq
(Strictly speaking, the observable charges have to be real and they
correspond therefore to the real and imaginary parts of ${\cal L}_{n,m}$).
One such charge is for example the angular momentum, which in the first
Landau level takes the form $J={\cal L}_{0,0}$.

When we switch on the confining potential $V_C$,
the Hamiltonian for the states in the first Landau level becomes
proportional to the angular momentum, $H_B+H_C=\alpha {\cal L}_{0,0}$,
see (\ref{20}).
Therefore, the charges no longer commute with the Hamiltonian,
\beq
{[H,{\cal L}_{n,m}]}=\alpha \ {[{\cal L}_{0,0},{\cal L}_{n,m}]}=
\alpha (n-m) \ {\cal L}_{n,m} \ .
\label{22}\eeq
They relate energy eigenstates to other energy eigenstates; thus
they give rise to a so-called {\it dynamical symmetry} \footnote{
Dynamical symmetries in $(2+1)$-dimensional non-relativistic physics
are also discussed in R. Jackiw's contribution to this volume.}
of the spectrum of the theory, whose properties we shall now discuss.

Note that the generators ${\cal L}_{nm}$ are defined in the full theory
and act ``horizontally'' within
each Landau level because they commute with the operators $d_i$ and
$d^{\dag }_i$. The commutator (\ref{22}) shows that the
${\cal L}_{n,m}$ are {\it raising} $(n >m)$ and {\it lowering }
$(n<m)$ operators for angular momentum and energy.

These generators satisfy the closed linear algebra
\beq
\begin{array}{l}
{\displaystyle {[{\cal L}_{n,m},{\cal L}_{k,l}]}=} \\
{\displaystyle \hbar \left( (m+1)(k+1)-
               (n+1)(l+1) \right) {\cal L}_{n+k,m+l}} \\
{\displaystyle + O({\hbar}^{2})\ {\cal L}_{n+k-1,m+l-1}}\\
{\displaystyle + O({\hbar}^3)+\dots\ ,}
\end{array}
\label{23}\eeq
where the missing terms correspond to contractions of more
derivatives and hence have higher powers of $\hbar$.
The complete algebra can be found in \cite{Cap1} and contains a finite
number of such terms.
If we truncate the r.h.s. of eq.(\ref{23}) to $O(\hbar)$, we obtain
the semiclassical limit of this algebra, which identifies it
as that of {\it area-preserving diffeomorphisms} or $w_\infty$.
The full algebra (\ref{23}) is the quantum version
called $W_{\infty}$ \cite{She1}.

Let us point out some
special cases of the full quantum algebra, besides (\ref{22}),
\barr
{[{\cal L}_{n,0} , {\cal L}_{k,0}]} &=& (k-n)\ {\cal L}_{n+k,0}\ ,\nonumber \\
{[{\cal L}_{0,n} , {\cal L}_{0,k}]} &=& (n-k)\ {\cal L}_{0,n+m}\ ,\nonumber \\
{[{\cal L}_{n,n} , {\cal L}_{k,k}]} &=& 0 \quad
   {\rm ( Cartan\ \  subalgebra)}\ .
\label{24}\earr
The first two sub-algebras are Virasoro algebras, with missing
negative $(n ,k< -1)$ modes. Actually,
$\ {{\cal L}}_{n,0} = \sum_{i} z_{i}^{n+1} \partial_{i}\ $,
when acting on analytic functions of the lowest level.
Note however that $[{\cal L}_{n0}, {\cal L}_{0k}]$ contains
several higher order in $\hbar$ terms; thus the two sub-algebras cannot
be combined into a single complete Virasoro algebra.
Nevertheless, we shall show in the next section how a full Virasoro algebra
with central charge $c=1$ emerges from $W_\infty$ when we describe
the dynamics of excitations at the edge of the incompressible
droplet in the thermodynamical limit.

It is instructive to illustrate the classical origin of the quantum
symmetry (\ref{23}). The Lagrangian for the theory truncated
to the first Landau level is obtained by taking the $m\to 0$
limit $(\omega\to\infty )$ of the Lagrangian corresponding to
\hbox{$H_B+H_C$\cite{Dun1} }:
\barr
L&=&\lim_{m\to 0} \left( L_B +L_C \right) \nonumber \\
&=& \sum_i \left(
\vec{ q}_i\wedge \dot{\vec q}_i + {1\over 2} \lambda \vert \vec q_i\vert^2
\right)\ .
\label{25}\earr
This Lagrangian is of first order in time derivatives; as a consequence
of the $m\to 0$ limit the original configuration space becomes a
$2^N$-dimensional {\it phase space} with Poisson brackets
$\{ \bar z_i,z_j \}_{PB}=i\delta_{ij}$.
Canonical transformations in the two-dimensional phase spaces are
{\it area-preserving diffeomorphisms;} they are dynamical symmetries
generated by ${\cal L}^{(cl)}_{n,m} =\sum_i z_i^{n+1} \bar{z}^{m+1}_i$
and satisfy the classical algebra $w_\infty$
\beq
\begin{array}{l}{\displaystyle
\{ {\cal L}^{(cl)}_{n,m},{\cal L}^{(cl)}_{k,l} \}_{PB}=} \\
{\displaystyle
\left( (m+1)(k+1)- (n+1)(l+1) \right) \ {\cal L}^{(cl)}_{n+k,m+l} \ .}
\end{array}
\label{26}\eeq
realized by Poisson brackets.

Having discussed the algebra satisfied by the generators ${\cal L}_{n,m}$,
we turn to the investigation of the action on the ground state $\psi_1$
in eq.(\ref{14}).
Since the ${\cal L}_{n,m}$ involve higher powers of the derivative,
they act non-locally on wave functions. The best strategy is to use
the second-quantized formalism, in which they are expressed in terms of
the field operators:
\beq
\begin{array}{l}{\displaystyle
{\cal L}_{n,m}
=\int d^2 \vec{x}\ \Psi^{\dag} c^{\dag n+1} c^{m+1}
\Psi =} \\
{\displaystyle
\sum_{k \ge m+1} {\sqrt{k! (k+n-m)!}\over{(k-m-1)!}}
a^{\dag}_{k+n-m} a_k \ .}
\end{array}
\label{27}\eeq
This representation makes it evident that the
ground state $\vert\Omega\rangle_1$
in eq.(\ref{15}) is annihilated by the lowering operators,
\beq
{\cal L}_{n,m} \ \vert\Omega\rangle_1\ =\ 0,\qquad n<m,\ m\ge 0\ ,
\label{28}\eeq
because the filled Landau level has the minimal angular momentum
(and energy) allowed by Fermi statistics.
In physical terms, these operators try to compress the fluid
by moving electrons to lower angular momentum states, but this is impossible
because all these states are already occupied. The conditions (\ref{28})
therefore express the {\it incompressibility} of the fluid.
Conditions like (\ref{28}) are called highest-weight conditions and precisely
formulate the {\it invariance} of the theory under the dynamical symmetry.
There are $N(N-1)/2$ non-trivial conditions for $N$
particles; in the thermodynamical limit $N\to\infty$ the incompressible
ground state is characterized by the highest-weight conditions of
the {\it infinite dynamical symmetry} $W_\infty$.

The raising operators ${\cal L}_{n,m}$ with $n>m$ applied on the ground
state generate all the excitations with wave functions
$\psi_{\Delta J}$ and $\Delta J=n-m$.
For $(n-m)=O(1)$ these are the gapless edge excitations;
for higher $(n-m)$ ($(n-m)=O(\sqrt{N})$ with our model harmonic $V_C$,
$(n-m)=O(N)$ in the realistic case) we obtain also gapful quasi-hole
excitations.
In other words, the situation is completely analogous to two-dimensional
conformal field theory: the ground state satisfies highest-weight
conditions while all excitations fall into representations of the
symmetry algebra \cite{Gin1}.

We now show that the $\nu=1/m $ incompressible Laughlin state $\psi_m$
satisfies similar infinite conditions.
As explained in section two, the interaction $H_I$ divides the
$N$-particle states of the first Landau level into two sub-bands,
eqs.(\ref{17},\ref{18}).
Let us call $\Theta$ the projector on the low-energy states of the
form (\ref{17}) and project the previous symmetry generators in this sub-band
\beq
{\hat{\cal L}}_{k,l} \equiv\ \Theta\ {\cal L}_{k,l}\ \Theta\ .
\label{29}\eeq
Their action amounts to compressions and decompressions of states
in this sub-band, transitions that can occur in the system due to
small perturbations.
It is therefore intuitive that the incompressible ground
state $\psi_m$ should be annihilated again by  the lowering operators,
\beq
{\hat{\cal L}}_{k,l} \ \vert\Omega\rangle_m\ =\ 0, \qquad k<l,\ l\ge 0\ .
\label{30}\eeq
Actually, this follows by the fact that the Laughlin ground state
is the state of lowest angular momentum in the low-energy sub-band,
as explained in \cite{Cap1}.
The algebra satisfied by the modified generators
${\hat{\cal L}}_{k,l}$ is still under investigation \cite{Cap3}.
Its classical limit is again the algebra of area-preserving
diffeomorphisms $w_\infty$ in eq.(\ref{26}).
Actually, the Laughlin states and the integer ground state are similar
in the classical limit, since they only differ by the value of the
constant density, and therefore have the same classical symmetry
$w_\infty$ \cite{Cap3}.
In the quantum case, the higher order terms $O(\hbar^n)$ in the algebra
$W_\infty$ might depend a priori on the filling $1/m$.

Our program is to characterize algebraically the
incompressibility of all quantum Hall ground states and thus to relate
them and their excitations to the representations of the $W_\infty$
algebra.
{}From our point of view this geometrical and algebraic
approach to the Quantum Hall effect is
most appropriate, as it can account for the exactness of the
conductivities and the observed universality.

\section{DYNAMICAL CONFORMAL SYMMETRY ON THE EDGE}

In this section we study the excitations at the edge of the quantum Hall
droplet.
We already saw that the generators ${\cal L}_{nm}$, for \hbox{$n-m=O(1)>0$},
give rise to small deformations of the droplet at constant density,
thus small edge deformations.
Moreover, Laughlin's quasi-particle excitations
\cite{Lau1} also affect the boundary.
An incompressible state is characterized by a constant density
$\rho$ in the bulk, whose total integral $\int d^2x\ \rho=N$
is also constant. Due to these properties, an excitation corresponding
to a localized density deformation is transmitted to the free boundary
(there are no density waves).

Since the bulk is rigid, the effective field theory of these edge
excitations lives in the $(1+1)$-dimensional spacetime given by
the boundary circle and time. We shall derive this effective theory from the
data of section two. We shall emphasize the property
of infinite conformal theory, which reflects the previous $W_\infty$-symmetry
in the bulk, and we shall show how it predicts {\it universal
finite-size terms} in the excitation spectrum.

Consider first the $\nu=1$ case of the lowest Landau level filled up to
(and including) angular momentum $L$. In section two we have discussed
the boundary interaction $H_C$ which lifts the degeneracy
of the Landau level. For $L\gg 1$ the one-particle spectrum
$\varepsilon_l$ can be taken as approximately linear in a region around $L$:
\beq
\varepsilon_l = \alpha\left( l-L \right) +\beta \ .
\label{epsilonell}\eeq
In eq.(\ref{20}) we showed that $\alpha$ has the finite-size expansion
\beq
\alpha={v\over R} + O\left({1\over R^2 }\right) \ ,
\label{alphaR}\eeq
where $R\simeq\sqrt{L}$ is the approximate radius of the droplet.
Taking into account the general finite-size expansion of $\beta$,
\beq
\beta =\beta_0 -{v\mu\over R} + O\left( {1\over R^2} \right) \ ,
\label{betaR}\eeq
the spectrum $\varepsilon_l$ up to order $O(1/R)$ takes the form
\beq
\varepsilon_l\ =\ \beta_0 + {v\over R}\left( l-L-\mu \right),
\label{epsilonR}\eeq
where $v$ is a parameter with the dimension of a velocity, whereas
$\mu$ is  dimensionless and plays the role of a chemical potential.
{}From the point of view of the edge there are two types of excitations:
neutral particle-hole pairs around the edge with energies of $O(1/R)$
and charged excitations, which amount to a transfer of particles
between the edge and the center of the droplet.
The former are the previously introduced gapless edge excitations \cite{Wen1}.
A positive charge on the edge corresponds to a quasi-hole in the bulk
and costs an energy $\beta_0 +O(1/R)$.
A negative charge on the edge corresponds to a quasi-particle in the bulk
and costs an energy $\Delta-\beta_0 +O(1/R)$, where $\Delta$ is the gap
to the next energy band (for $\nu=1$ we have e.g. $\Delta=\omega$).
These are the Laughlin quasi-particle excitations.

The implications of a linear spectrum near the edge are of great significance
only if this linear range expands to infinity in the thermodynamical
limit $L\to\infty$.
To study this situation in more detail, we apply the confining pressure
to the electrons by a novel procedure. Instead of introducing
an harmonic confining potential,
we pick a circle of radius $R$ in the infinite plane and modify the
(second-quantized) Hamiltonian (\ref{8}) to
\beq
H_{R} = {1\over 2} \int_{R} d^{2} {\vec x} \left[
{\left ( \vec D \Psi \right )}^{\dag} {\left( \vec D \Psi \right)}
- 2 \rho \right] \ ,
\label{bounh}\eeq
where the subscript $R$ denotes integration of the radial
coordinate only up to $R$ \footnote{We remind the reader that we have set
$M=1, B=2, e=1$.}.
While being defined in the infinite plane, the new Hamiltonian mimics
the system confined to a disk of radius $R$, by
relating the new parameter $R$ to the particle number $L+1$ via
\beq
R^{2} = L+\mu\ ,\qquad L\gg 1 \ ,\ \mu = O(1)\ .
\label{defmu}\eeq
The eigenfunctions $\psi_{l}$ (see eq.(\ref{7}))
of the first Landau level still diagonalize $H_{R}$ (the angular integration
is the same).
The first Landau level Fock space is thus unchanged; however
the degeneracy is lifted and we obtain the one-particle spectrum
\beq
\varepsilon_{l}={R^{2l}\ \exp (-R^2)\over{l!}} \ (l-R^2 )\ .
\label{espect}\eeq
As a consequence, the field operator restricted to the first Landau level
acquires a time dependence given by
\beq
\Psi({\vec x},t) = \sum_{l=0}^{\infty}\ a_{l}\
\psi_{l}({\vec x}) \ e^{-i\varepsilon_{l} t} \ .
\label{flltd}\eeq

For $L-\sqrt{L} \le l\le L+\sqrt{L}$, the spectrum $\varepsilon_{l}$
becomes approximately linear
\beq
\varepsilon_{l}\simeq {v\over R}\left( l-L - \mu \right)\ ,\qquad
v={1\over\sqrt{2\pi}} \ ,
\label{espcm}\eeq
and reproduces the realistic spectrum (\ref{epsilonR}) with
the only exception that $\beta_0 =0$.
For the moment we consider this case $\beta_0=0$; we shall
easily restore the important parameter $\beta_0$ a posteriori.
Note also that the velocity parameter $v$ is to be considered
as a {\it phenomenological parameter}; its exact value depends
on the real confining potential felt by the electrons and
the value in (\ref{espcm}) is not to be taken as a prediction.
Furthermore, the parameter $\mu$ will be later determined dynamically.

The spectrum (\ref{espcm}) indicates that the edge excitations
are governed by a {\it relativistic (1+1)-dimensional theory}
with effective ``light-velocity'' $v$.
This is indeed the case; the Hamiltonian for this relativistic theory
can be found easily by using the identity
\barr
\left(\vec D\Psi\right)^{\dag} \left(\vec D \Psi\right)&=&
\left( D_{+}\Psi\right)^{\dag} \left(D_{+}\Psi\right) \nonumber\\
&+& \vec\nabla\wedge\vec J + 2\rho
\label{bogide}\earr
where $D_{+}=D_x +i D_y$ and $\vec J$ is the non-relativistic
current density
\beq
\vec J={1\over 2i}\left\{ \Psi^{\dag}\vec D\Psi -
\left( \vec D\Psi\right)^{\dag} \Psi \right\}.
\label{curre}\eeq
Given that the covariant derivative $D_{+}$ is proportional to the
operator $d$ in eqs.(\ref{5}),(\ref{6}) and annihilates the first
Landau level field operator (\ref{flltd}), we recognize that $H_{R}$ reduces to
a pure boundary term
\beq
H_{R}\ =\ {1\over{4}} \int_{0}^{2\pi R} dx\
{\Psi}^{\dag}\left(-i\partial_{x} - R\right) \Psi + c.c.\ ,
\label{boham}\eeq
where $x=R\theta$ is the one-dimensional coordinate on the edge at $r=R$.

To proceed further, it is convenient to shift variables as follows:
\barr
e_{l} &\equiv & \varepsilon_{L+l}\ ,\nonumber\\
b_{l} &\equiv & a_{L+l}\ , \label{shift}\\
\Psi ({\vec x}) &= & \left( {2\over \pi} \right)^{1/4}\
e^{i(L+\mu)\theta} \ F_{R}^{\rm (C)} ({\vec x})\ .\nonumber
\earr
When evaluated on the boundary, the field operator
$F_R^{\rm (C)}$ acquires the form
\beq\begin{array}{l}
{\displaystyle
F_R^{\rm (C)} \left(Re^{i\theta}, t \right)
= \sum_{l=-L}^{\infty} {C_l\over\sqrt{2 \pi R}}
e^{i(l-\mu) \theta } e^{-i e_l t}\ , }\\
{\displaystyle
C^2_l = \left. {\sqrt{2\pi}\over{(L+l)!}}
 R^{2L+2l+1}  e^{-R^2}\right\vert_{R^2 =L+\mu} \ .}
\end{array}\label{expf}\eeq

In this expression, the coefficients are approximated by
$C_l\sim \exp(-(l-\mu)^2/2L)$ for $L\to\infty$.
Since the width of this Gaussian is $\sqrt{L}$, the $C_l$
give a smooth ultraviolet cutoff to the sum over $l$, which is limited
to the range $|l|<\sqrt{L}\simeq R$ of linear
energy (\ref{espcm}). In this range, the wave functions have the correct
$(1+1)$-dimensional form which ensures orthonormality in the
Hilbert space on the circle.
Note that to leading order for $L\to\infty$, we can remove
this cutoff which would give subleading contributions
to $O(1/R)$.
Therefore, we obtain the approximate field operator
\beq
F_R \left(x, t \right)
= \sum_{l=-\infty}^{+\infty} {1\over{\sqrt{2 \pi R}}}\
e^{i(l-\mu) (x-vt)/R}\ .
\label{feff}\eeq

In terms of the field $F_{R}$, the Hamiltonian
$H_R$ in eq.(\ref{boham}) takes the form
\barr
H_{R}& =& {v\over 2} \int_{0}^{2\pi R} dx\
F^{\dag}_{R} \left( -i\partial_{x} \right) F_{R}
+ c.c.\ .
\label{boshi}\earr

Equations (\ref{feff})  and (\ref{boshi})  are the field and the
Hamiltonian of a relativistic, chiral, charged
fermion (Weyl fermion), moving with ``light-velocity'' $v$ on a circle of
circumference $2\pi R$. Note that the filled Landau level plays the
role of the {\it Dirac sea} for this relativistic fermion \cite{Sto1}.
The fermion field (\ref{feff}) has a general boundary condition
\beq
F_R(\theta +2\pi) =\  e^{-i2\pi\mu} \ F_R(\theta)\ ,
\label{boundary}\eeq
parametrized by the fractional part of $\mu$.
Such boundary conditions are possible for the Weyl fermion
\cite{Gin1}, and affect the spectrum of the theory, as we shall see
later. The two cases mostly considered in  string theory are
called:
{\it i)} $\mu=0$: Ramond (R),  or periodic b.c. on the circle;
{\it ii)} $\mu={1\over 2}$: Neveu-Schwarz (NS), or anti-periodic
b.c. on the circle.

\begin{figure*}[t]
  \vspace{9cm}
  \caption{The droplet of electrons in the physical $z$-plane and the
space-time cylinder $(x=R\theta,t)$ where the conformal field theory of edge
excitations lives.}
\end{figure*}

Let us now show that this theory of a Weyl fermion
is a {\it conformal field theory} defined
on the {\it cylinder} made by the circle and Minkowskian time $t$
(see Figure 2). Define the charges
\barr
\rho_n & = &\int_0^{2\pi R} dx\ F_R^{\dag}F_R \ \ e^{-in(x-vt)/R}\ ,
\nonumber\\
L_n & = & {R\over v} \int_0^{2\pi R} dx \ {\cal H}_R \ \ e^{-in(x-vt)/R}\ ,
\label{curr}\earr
where ${\cal H}_R$ is the Hamiltonian density in (\ref{boshi}).
$\rho_n$ and $L_n$ generate local gauge and conformal transformations
respectively and can be written in terms of the Fock space operators as
\barr
{\rho}_{n} &=& \sum_{l=-\infty}^{+\infty}
\ b^{\dag}_{l-n}b_{l}\ , \nonumber\\
L_{n} &=& \sum_{l=-\infty}^{+\infty}\ (l-{n\over 2} -\mu)
\ b^{\dag}_{l-n}b_{l}\ .
\label{modes}\earr
Clearly, $\rho_0$ and $L_0$ have to be
regularized with some normal ordering prescription.
This amounts to an infinite subtraction achieved by the standard procedure
of writing annihilators to the right-hand side of creators,
plus possible additional finite normal ordering constants.
These are fixed by algebraic conventions as explained e.g. in
\cite{Itz2}.

The normal-ordered charges $\rho_n$ and $L_n$ satisfy the following
chiral algebra
\barr
[{\rho}_{n},{\rho}_{m}] &=& n\ \delta_{n+m,0} \ , \nonumber\\
{[{ L_{n},{\rho}_{m} }]} &=& -m\ \rho_{n+m}\ , \nonumber\\
{[{ L_{n},L_{m} }]} &=& (n-m)L_{n+m} \nonumber \\
&+& {c\over 12}(n^{3} - n)
\delta_{n+m,0}\ ,\quad c=1\ .
\label{chalg}\earr
The first algebra in (\ref{chalg}) is a {\it U(1) Kac-Moody algebra}
while the third algebra is a {\it Virasoro algebra}.
As is well known, the central extensions in
these two algebras come from quantum effects associated with the
infinite Dirac sea in the thermodynamic limit. The coefficient of $n^3$
in the Virasoro algebra is independent of finite normal-ordering
constants and fixes the central charge to be $c=1$,
as expected for a Weyl fermion.
The coefficient of the linear term in the
Virasoro central extension can be shifted by redefinitions of
$L_0$ by a constant. The convention in (\ref{chalg}) is standard and
fixes the value of $L_0 |\Omega\rangle$. The value of
$\rho_0 |\Omega\rangle$ is fixed by the requirement of the
absence of an anomaly in the mixed commutator.
The properties of the ground state are summarized by
\barr
L_n |\Omega,\mu\rangle   &=& 0\ ,\quad
     \rho_n |\Omega,\mu\rangle = 0 \ ,\quad  n>0\ , \nonumber\\
L_{0}|\Omega,\mu\rangle  &=& \left({1\over 8}+{{\mu^{2}-\mu}\over 2}
     \right) |\Omega,\mu\rangle\ ,\nonumber\\
\rho_0|\Omega,\mu\rangle &=&
     \left( {1\over 2} - \mu \right) |\Omega,\mu\rangle\ .
\label{vacpr}\earr

In the thermodynamic limit, the filled Landau level is a highest-weight
state of the ($c=1$) chiral algebra (\ref{chalg}) with charge
$Q_0={1\over 2} -\mu$ and conformal dimension
$h_0={1\over 8}+{1\over 2}({{\mu}^{2}-\mu})$.
Given that $|\Omega,\mu\rangle$ represents the original filled Landau
level state, we recognize (\ref{vacpr}) as an {\it algebraic
characterization of the incompressibility} of this state.
It can be shown \cite{Cap3}
that the highest-weight conditions (\ref{vacpr}) are a subset of
the $W_\infty$-symmetry conditions discussed in section three.

Note that $\rho_0$ is the renormalized charge operator, which correctly
couples to additional electro-magnetic fields, thereby leading to the measured
Hall current. Therefore, the ground state should be a neutral eigenstate
of $\rho_0$. We thus conclude that $\mu$ dynamically self-tunes
to the value
\beq
\mu\ =\ {1\over 2}\ .
\label{muns}\eeq
This implies that the ground state
$|\Omega\rangle \equiv |\Omega,{\mu ={1\over 2}}\rangle$
is $SL(2,{\bf C})$ invariant, since it is also annihilated by all three
generators of global conformal transformations,
$ L_{k}|\Omega\rangle =0\ , \ k=-1,0,1\ $.
We thus conclude that the theory describing the dynamics of the
edge excitations in the thermodynamic limit is the $c=1$ CFT
of a Weyl fermion with Neveu-Schwarz boundary conditions, eq.(\ref{boundary}).

We now have to discuss the normal ordered Hamiltonian. From
(\ref{curr}) we would naively conclude that $H_R=vL_0/R$.
However, the normal ordering leading to (\ref{chalg}) and (\ref{vacpr})
implies a $c$-number correction \cite{Car1}
\beq
H_R\ =\ {v\over R}\left(L_0 -{c\over 24}\right)\ , \qquad (c=1)\ ,
\label{casen}\eeq
due to the {\it conformal anomaly}.
The additional term, depending on the central
charge $c$, is the {\it Casimir energy} to be discussed below.

\section{FINITE SIZE EFFECTS}

\noindent{\bf 5.1. Integer filling}

Having derived the symmetry algebra and the Hamiltonian, we are ready
to discuss the operator content of the theory. Indeed, all excitations
above $\vert\Omega\rangle$ can be classified according to irreducible
highest-weight representations of the infinite chiral algebra (\ref{chalg}).
As is well known \cite{Gin1}, these are encoded
in the grand-canonical partition
function. Before computing the partition function let us
however restore a non-vanishing $\beta_0$ in (\ref{epsilonR}).
The crucial point is that $\rho_0$ is a Casimir of the chiral algebra
(\ref{chalg}). Therefore, one can modify the Hamiltonian by adding to
it any function of $\rho_0$, without changing the Hilbert space
of the theory. We can thus add gaps proportional to the charge
of excitations with the following redefinition:
\barr
H_R \ \to {\hat H}_R & =& {v\over R}
\left( L_0 -{1\over 24} \right) +\gamma (\rho_0) \ ,\nonumber\\
\gamma (\lambda ) &=& \left\{
    \begin{array}{ll}
    \lambda \beta_0,             &  \lambda \ge 0\ , \\
    \lambda (\Delta - \beta_0 ), &  \lambda <0 \ ,
    \end{array}\right.
\label{hhat}\earr
where $\lambda$ labels the eigenvalues of $\rho_0$.
In any irreducible representation with charge $\lambda$ the additional
piece is a multiple of the identity; thus it still makes sense to label
states of the Hilbert space according to representations of (\ref{chalg}).
The resulting theory is a fully consistent {\it deformed } CFT.

The grand-canonical partition function
\barr
{\cal Z}(q,w) &=&
Tr \ q^{L_0 -{ 1\over 24} + {R\over v}\gamma (\rho_0) } \
w^{\rho_0} \ ,\nonumber\\
q &\equiv & e^{-\beta v/R}, \quad
w=\ fugacity \ ,
\label{grandpar}\earr
can be computed with standard techniques. The result is \cite{Cap2}
\beq
{\cal Z}(q,w) = {1\over \eta (q)} \
\sum_{k=-\infty}^{+\infty}
q^{ {1\over 2}k^2 + {R\over v}\gamma (k)} \ \ w^k \ ,
\label{zdisk} \eeq
with $\eta (q) = q^{1/24} \prod_{n=1}^{\infty} (1-q^n)$
the Dedekind function.

Given this result we can discuss the whole spectrum and the finite size
effects.
The {\it ground state energy} $E_0$ will in general have a finite-size
expansion in powers of $R$ :
\beq
E_0 =\alpha_2 R^2 +\alpha_1 R^1 +\alpha_0 + \alpha_{-1} {1\over R} +
\alpha_{-2} {1\over R^2} +\dots \ .
\label{ezero}\eeq
The leading term gives a finite energy density, while the remaining terms
are subleading in the experimental regime of large flux through the sample
($\Phi= R^2 \gg 1$). On the other hand, the result of CFT is given by
eq.(\ref{casen}) combined with
$\rho_0|\Omega\rangle= L_0|\Omega\rangle= 0$:
\beq
E_0 = - {v\over 24} {1\over R} \ .
\label{eground}\eeq
In CFT the terms $\alpha_2$ , $\alpha_1$ and $\alpha_0$ have been put
to zero by the infinite subtraction of normal ordering.
The terms $\alpha_{-2},\alpha_{-3},\dots$ instead, represent the
higher order corrections neglected in thermodynamic limit $R\to\infty$.
Thus, CFT leads to the prediction $\alpha_{-1} = -{v/24}$;
this prediction is universal because it does not depend on the details
of the confining potential other than the velocity $v$.
Moreover, this unique parameter can be measured independently from the
spectrum of excitations, as we now discuss.

The partition function (\ref{zdisk}) contains two basic factors. The factor
$1/\eta (q)$ represents the contribution from neutral particle-hole
excitations across the Fermi level; the second factor is a sum
over charged sectors which corresponds to quasi-holes and quasi-particles
in the  bulk. The charge of these excitations is given by
\beq
Q_k= k\ .
\label{chargeqp}\eeq
The total spectrum of excitations can be read off (\ref{zdisk}):
\beq
\Delta E_{k,n}\equiv E_{k,n} - E_0 = \gamma (k)  + {v \over R}{k^2\over 2}+
{vn\over R}\ ,
\label{qhqp}\eeq
with $k,n \in {\bf Z}$ and $0<n\ll R$ .
The first term $\gamma (k)$ describes the gap for an excitation of
charge $k$. This term is non-universal and is not predicted by CFT.
The second term is the {\it universal $1/R$-correction}
to the gap; CFT predicts that this is tied to the charge $k$ and
has the same value for quasi-particles and quasi-holes.
Its value is the conformal dimension $h_k =k^2 /2$, the eigenvalue
of $L_0$ which labels the irreducible representations of the CFT.
The third term is the spectrum of neutral edge excitations
in any given charged sector (the conformal family).

\bigskip
\noindent{\bf 5.2. Fractional filling}
\bigskip

Up to now we have discussed only the case of filling fraction $\nu=1$.
For the Laughlin fluids with $\nu=1/m$ we do not have a direct
derivation of the CFT in terms of the original fermionic variables.
However, our symmetry arguments of section three indicate that the
universal, long-range properties of these fractional-filling fluids are again
described by a $c=1$ CFT. Indeed, it can be shown that the highest-weight
conditions (\ref{vacpr}) are a consequence of the $W_\infty$ highest-weight
conditions (\ref{28}). Having shown that the $\nu=1/m$ states have the
same symmetry as the $\nu=1$ one, it seems very plausible that they
correspond to different sets of representations of the $c=1$ chiral
algebra (\ref{chalg}).
Clearly, the $U(1)$ symmetry associated to charged excitations is still
present and requires that the CFT has $c=1$ (or many copies of $c=1$ CFTs).
Moreover, the form of Laughlin's quasi-particle wave functions for $\nu=1/m$
\cite{Lau1} is very reminiscent of correlators of a $c=1$ CFT \cite{Fub1}.

There is a one-parameter family of $c=1$ CFTs giving
all representations of the chiral algebra (\ref{chalg}) \cite{Gin1}.
These theories describe an interacting fermion with varying coupling constant.
Therefore the previous free fermionic picture ($\nu=1$) cannot be
easily extended to a generic $c=1$ CFT. Fortunately, the general theory for
the interacting fermion is equivalent, at the quantum level, to the
theory of a free {\it chiral boson} \cite{Gin1}.
The convenient action for the chiral boson is \cite{Flo1} \cite{Cap2}:
\beq
S =-{\kappa\over{4 \pi}} \int_{-\infty}^{+\infty}
dt \int_{0}^{2\pi R} dx \left({\partial}_{t} + v {\partial}_{x}
\right)\phi {\partial}_{x} \phi\ .
\label{boso}\eeq
The coupling $\kappa$ allows to span the $c=1$ CFTs and can only take
{\it rational } values for a consistent quantization of the theory \cite{Cap2}.
Its values for the Laughlin states,
\beq
\kappa={1\over \nu}=m\ ,
\label{kappa}\eeq
are identified as follows \cite{Cap2}.

The chiral boson theory has a {\it chiral anomaly}, similarly to the previous
case of the Weyl fermion theory \cite{Tre1}. This implies that the charge is
not
conserved when the theory is coupled to an external electromagnetic field,
\beq
\partial_t Q={e\over \kappa} E \ ,
\label{anoma}\eeq
where $E$ is the electric field on the edge.
This equation tells us the rate of production of particles
by the external electric field.
{}From the $(1+1)$-dimensional point of view, this charge pops out from the
Dirac sea (spectral flow picture), which is an infinite reservoir.
The identification of the Dirac sea with the quantum
Hall droplet allows then to interpret the anomaly as a radial flow
of charge through the boundary at $R$, {\it i.e.}, a Hall current
\beq
{J^{i}\ =\nu {e\over{2\pi}}\ \epsilon^{ij} E^{j}\ .}
\label{hallcu}\eeq

Let us describe the results of this identification \cite{Cap2}.
The finite-size Casimir energy $E_0=-v/24 R$
is independent of $m$. The {\it fractional charges} of quasi-holes and
quasi-particles are given by
\beq
Q_k ={k\over m}\ , \qquad (k\  integer)\ ,
\label{frachar}\eeq
which matches perfectly Laughlin's result. The excitation spectrum
is found to be
\beq
\Delta E_{k,n}=\gamma (k) + {v \over R}{k^2\over 2m} + {v n\over R}\ .
\label{fracR}\eeq
As in eq.(\ref{qhqp}), the first two terms in this equations are
the non-predictable, non-universal gaps and their predicted universal
$O(1/R)$ corrections.
The third term gives the neutral, gapless excitations with spectrum $vn/R$.
Their multiplicities are the same as in the $\nu=1$ case, since they
can be deduced from the Dedekind function in (\ref{zdisk}).

These multiplicities have indeed been confirmed by numerical calculations
of {\it interacting} electrons on a disk geometry \cite{Wen1} \cite{Sto2}.
This is a first non-trivial verification of the CFT description
of edge excitations.

Our main contribution here is the quantitative prediction
(\ref{fracR}) for the $O(1/R)$ dependence of the gaps.
Note that the finite-size scaling of the neutral excitations
allows the identification of $v$. The finite-size scaling of the
ground state energy and the gaps is then completely fixed by (\ref{fracR}).
This would provide a new and different confirmation of Laughlin's
theory and show that this is, in a sense, ``equivalent'' to a $c=1$ CFT.
A numerical verification of our results is difficult, since one needs
Monte-Carlo calculations for large number of electrons and the correct
inclusion of all interactions. The existing calculations \cite{Morf}
are unfortunately not accurate enough for extracting the $(1/R)$ dependence.

The {\it fractional statistics} of quasi-particle excitations
can also be accounted for by this chiral CFT, as a result of its
conformal dimensions and charges and the abelian monodromies
of $c=1$ correlators. We address the reader to our paper \cite{Cap2} for
a complete discussion. Let us only mention a simple consequence.
Having identified the CFT on the boundary circle of the quantum Hall droplet,
we can relate CFT correlators to correlators and wave function of the
physical problem in $(2+1)$ dimensions. Therefore, we can physically justify
the formal relations between CFT and quantum Hall effect discussed in the
refs.\cite{Fub1}.

\section{CONCLUSIONS}

Laughlin's fundamental physical picture of {\it incompressible
quantum fluids} has been related to a symmetry principle.
This is likely to apply to all the fractional Hall plateaus of the
hierarchical construction \cite{Hal1}\cite{Jai1}, which presumably
correspond to different irreducible representations of the
$W_{\infty}$ algebra.
More work is necessary to develop the necessary mathematical machinery.

The study of edge excitations has been already very effective,
because the infinite conformal symmetry is much better understood.
By developing known results on the chiral fermion and its bosonization,
we obtained the charges and conformal dimensions of quasi-particle
excitations.
Conformal field theory gives an {\it a posteriori justification }
of Laughlin's theory of quasi-particles excitations, including their exact
fractional statistics, and of the exact fractional Hall conductivity
(see eqs.(\ref{kappa},\ref{hallcu})).
Moreover, the CFT prediction on the finite-size energy spectrum
(\ref{fracR}) gives a {\it direct} way to {\it verify} Laughlin's
theory in numerical experiments of the electron fluid.


\begin{thebibliography}{9}
\bibitem{Pra1} For a review see: R. A. Prange, S. M. Girvin, "The Quantum
               Hall Effect", Springer Verlag, New York (1990).
\bibitem{Kino} M. E. Cage et al., IEEE Trans. Instrum. Meas. 38 (1989) 284;
               T. Kinoshita, IEEE Trans. Instrum. Meas. 38 (1989) 172.
\bibitem{Lau1} For a review see: R. B. Laughlin, "Elementary Theory: the
               Incompressible Quantum Fluid", in \cite{Pra1}.
\bibitem{Hal1} For a review see: F. D. M. Haldane, "The Hierarchy of
               Fractional States and Numerical Studies", in \cite{Pra1}.
\bibitem{Jai1} For a review see: J. K. Jain, Adv. in Phys. 41 (1992) 105.
\bibitem{Gin1} For a review see: P. Ginsparg, "Applied Conformal Field Theory",
               in "Fields, Strings and Critical Phenomena", Les Houches 88,
               E. Brezin and J. Zinn-Justin eds., North-Holland, Amsterdam
               (1990).
\bibitem{Fub1} S. Fubini, Mod. Phys. Lett. A6 (1991) 1779; S. Fubini and
               A. L\"utken, Mod. Phys. Lett. A6 (1991) 487; G. V. Dunne,
               A. Lerda and C. A. Trugenberger, Mod. Phys. Lett. A6 (1991)
               2819; C. Cristofano, G. Maiella, R. Musto and F. Nicodemi,
               Phys. Lett. B262 (1991) 88, Mod. Phys. Lett. A6 (1991) 1779,
               Mod. Phys. Lett. A6 (1991) 2985; G. Moore and N. Read, Nucl.
               Phys. B360 (1991) 362.
\bibitem{Fro1} J. Fr\"ohlich and T. Kerler, Nucl. Phys. B354 (1991) 369.
\bibitem{Cap1} A. Cappelli, C. A. Trugenberger and G. R. Zemba, "Infinite
               Symmetry in the Quantum Hall Effect", CERN preprint
               CERN-TH 6516/92, to appear in Nucl. Phys. B.
\bibitem{Cap2} A. Cappelli, G. V. Dunne, C. A. Trugenberger and G. R. Zemba,
               "Conformal Symmetry and Universal Properties of Quantum Hall
               States", CERN preprint CERN-TH 6702/92.
\bibitem{She1} For a review see: X. Shen, "W-Infinity and String Theory",
               CERN preprint CERN-TH 6404/92.
\bibitem{Tru1} S. A. Trugman and S. Kivelson, Phys. Rev. B31 (1985) 5280.
\bibitem{Gir1} S. M. Girvin and T. Jach, Phys. Rev. B29 (1983) 5617;
               C. Itzykson, "Interacting Electrons in a Magnetic Field", in
               "Quantum Field Theory and Quantum Statistics, Essays in
               Honor of the 60th Birthday of E. S. Fradkin", A. Hilger,
               Bristol (1986).
\bibitem{Halp} B. I. Halperin, Phys. Rev. B25 (1982) 2185.
\bibitem{Wen1} For a review see: X.-G. Wen, Int. Jour. Mod. Phys. B6
               (1992) 1711.
\bibitem{Pra3} For a review see: R. A. Prange, ``Effects of imperfections
               and disorder'', in \cite{Pra1}.
\bibitem{Prui} A. M. M. Pruisken, ``Field Theory, Scaling and the Localization
               Problem'', in \cite{Pra1}; C. A. L\"utken and G. G. Ross,
               Phys. Rev. B45 (1992) 11837.
\bibitem{Dun1} G. V. Dunne, R. Jackiw and C. A. Trugenberger, Phys. Rev.
               D41 (1990) 661; G. V. Dunne and R. Jackiw, MIT preprint
               CTP 2123 (July 92), to appear in Umezawa volume.
\bibitem{Cap3} A. Cappelli, C. A. Trugenberger and G. R. Zemba, CERN preprint,
               in preparation.
\bibitem{Sto1} M. Stone, Ann. Phys. (NY) 207 (1991) 38, Phys. Rev. B42
               (1990) 8399, Int. Jour. Mod. Phys. B5 (1991) 509.
\bibitem{Itz2} C. Itzykson and J.-M. Drouffe, ``Statistical Field Theory'',
               Cambridge Univ. Press, Cambridge, (1988).
\bibitem{Car1} For a review see: J. L. Cardy, "Conformal Invariance and
               Statistical Mechanics", in "Fields, Strings, and Critical
               Phenomena", Les Houches 88, E. Brezin and J. Zinn-Justin
               eds., North-Holland, Amsterdam (1990).
\bibitem{Flo1} R. Floreanini and R. Jackiw, Phys. Rev. Lett. 59 (1987) 1873.
\bibitem{Tre1} For a review see: S. B. Treiman, R. Jackiw, B. Zumino and
               E. Witten, "Current Algebra and Anomalies", Princeton
               University Press, Princeton NJ (1985).
\bibitem{Sto2} M. Stone, H. W. Wyld and R. L. Schult, "Edge Waves in the
               Quantum Hall Effect and Quantum Dots", preprint ILL(TH) 91-27.
\bibitem{Morf} See for example: R. Morf and B. I. Halperin, Phys. Rev. B33
               (1986) 2221.
\end{thebibliography}
\end{document}